\documentclass[3p,times]{elsarticle}

\usepackage{ecrc}

\pdfoutput=1

\volume{00}

\firstpage{1}

\journalname{Nuclear Physics A}

\runauth{}

\jid{nupha}

\usepackage{amssymb}

\usepackage[figuresright]{rotating}

\begin{document}

\begin{frontmatter}




\title{Heavy flavor production in pp and AA collisions at the LHC}


\author[DFUNITO,INFNTO]{W.M.~Alberico}
\author[CERN]{A.~Beraudo}
\author[INFNTO]{A.~De Pace}
\author[INFNTO]{M.~Monteno\corref{cor1}}
\ead{monteno@to.infn.it}
\author[DFUNITO,INFNTO]{A.~Molinari}
\author[INFNTO]{M.~Nardi}
\author[INFNTO]{F.~Prino}

\cortext[cor1]{Corresponding author}

\address[DFUNITO]{Dipartimento di Fisica dell'Universit\`a di Torino, via P.~Giuria 1, I-10125 Torino, Italy}
\address[INFNTO]{Istituto Nazionale di Fisica Nucleare, Sezione di Torino, via P.~Giuria 1, I-10125 Torino, Italy} 
\address[CERN]{Physics Department, Theory Unit, CERN, CH-1211 Gen\`eve 23, Switzerland}

\begin{abstract}
A refined version of a multi-step calculation of heavy-flavor observables 
in pp and AA collisions has been developed, based on pQCD at NLO accuracy 
followed by parton shower evolution to describe heavy-quark production and 
on the relativistic Langevin equation to describe their stochastic evolution 
in the QCD plasma. Then, hadronization is modeled through an implementation of 
fragmentation functions based on pQCD and constrained by $e^{+}e^{-}$ collider data.
Results of our calculations can be compared with recent measurements 
performed at the LHC in Pb--Pb collisions at $\sqrt{s_{NN}}$=2.76~TeV:
nuclear modification factor $R_{AA}$ of the $p_{T}$ spectra at mid-rapidity 
of heavy-flavor decay electrons and of exclusively reconstructed open-charm 
mesons at different centralities, as well as their elliptic-flow $v_{2}$($p_{T}$) 
in semi-central collisions.
To test the validity of our setup for such studies, its predictions are 
also checked against the $p_{T}$ spectra measured in pp collisions at 
$\sqrt{s}$=7~TeV and 2.76~TeV.
\end{abstract}


\end{frontmatter}



\section{Introduction}

Heavy-flavor measurements in nucleus--nucleus collisions allow to test 
theoretical predictions about partonic energy-loss in the hot and ultra-dense 
strongly-interacting matter that is produced.

Results of the heavy-flavor measurements performed by PHENIX and STAR at the 
RHIC have shown that the suppression of the heavy-flavor decay electron 
$p_{T}$ spectra observed in central Au--Au collisions at $\sqrt{s_{NN}}$=200~GeV
is underestimated by model calculations implying for heavy-quarks only 
in-medium radiative energy losses. Such scenario has been confirmed by the 
first results delivered after the analysis of data collected in Pb--Pb 
collisions at $\sqrt{s_{NN}}$=2.76~TeV by the ALICE experiment at the LHC, where 
the exclusive reconstruction of open-charm hadrons from their hadronic decays
complements the inclusive measurement of electron (and muon) spectra.

Such findings gave boost to calculations considering the role of collisional 
energy loss for heavy quarks.

\section{The POWLANG model}

In our framework~\cite{Langevin2011,QM2011} the heavy-quark propagation in QGP is
described through a relativistic Langevin equation, where the stochastic noise term 
is expressed in terms of two momentum-dependent transport-coefficients representing 
the mean squared transverse/longitudinal momentum-change per unit time of the scattered 
heavy quark.
They are computed through pQCD calculations with resummation of medium effects 
(in HTL approximation) for soft collisions. 

The Langevin simulation tool is embedded in a multi-step setup to calculate 
heavy-flavor observables in $pp$ and AA collisions: 
$c$ and $b$ quarks are generated using the POWHEG~\cite{POWHEG-hvq} 
code, implementing pQCD at NLO accuracy, with CTEQ6M PDFs as input, followed  
by parton shower evolution performed with the PYTHIA code~\cite{Pythia} 
(this feature was not present in the previous version of our tool 
~\cite{Langevin2011,QM2011}. For AA collisions, EPS09~\cite{EPS09} nuclear 
corrections to PDFs are employed; then, heavy quarks are distributed in 
the transverse plane according to the nuclear overlap function 
$T_{AB}(x,y)\!\equiv\!T_A(x\!+\!b/2,y)T_B(x\!-\!b/2,y)$ corresponding to the 
selected impact parameter $b$; a broadening of the heavy-quark $p_T$-spectra 
due to an intrinsic-$k_T$ (in pp) and to the Cronin effect (in AA) is also 
included. 
 
Then, only for AA collisions, an iterative procedure is employed to follow 
the stochastic evolution of the heavy quarks in the plasma until hadronization: 
the Langevin update of the heavy-quark momentum and position is performed 
at each step according to the local 4-velocity and temperature $T(x)$ of 
the expanding background medium, as provided by relativistic hydrodynamic 
codes~\cite{kolb,romatschke}.

Heavy quarks are made hadronize by sampling different hadron species   
from $c$ and $b$ fragmentation fractions extracted from experimental data
~\cite{ZEUS,ALEPH1, HFAG}; this approach neglects, by construction, any 
possible change in the heavy-flavor hadro-chemistry in AA collisions. Then, 
hadron momenta are sampled from fragmentation functions (FF) calculated in 
heavy-quark effective theory (HQET)~\cite{Braaten}, with a single parameter 
$r$ defined as $r=(m_{H}-m_{Q})/m_{H}$. In order to evaluate the systematic
uncertainties associated to different FF choices, for charmed hadrons we have 
tested also the values of the $r$ parameter fitted in the FONLL framework 
~\cite{FONLL_D_Tevatron} to ALEPH data~\cite{ALEPH1} at the LEP $e^{+}e^{-}$ collider 
(i.e. $r=0.1$ for $m_{c}=1.5$~GeV, while it amounts to $r$=0.2 according its 
definition in~\cite{Braaten}; for bottom fragmentation we tested the functional 
form proposed by Kartvelishvili et al.~\cite{KAR}, whose single parameter 
$\alpha$ was fitted in the FONLL framework\cite{FONLL_B_Tevatron} to ALEPH~\cite{ALEPH2} 
and SLD~\cite{SLD} $e^{+}e^{-}$ data (namely $\alpha=29.1$ for $m_{c}=4.75$~GeV). 

Finally, each heavy-quark hadron is forced to decay into electrons with 
PYTHIA~\cite{Pythia}, using updated tables of branching ratios based 
on Ref.~\cite{PDG2012}. 

For brevity, in the following our setup will be named POWLANG
(POWHEG+Langevin). 

\section{Results}

\begin{figure}[htb]
\begin{center}
\includegraphics[clip,width=0.48\textwidth]{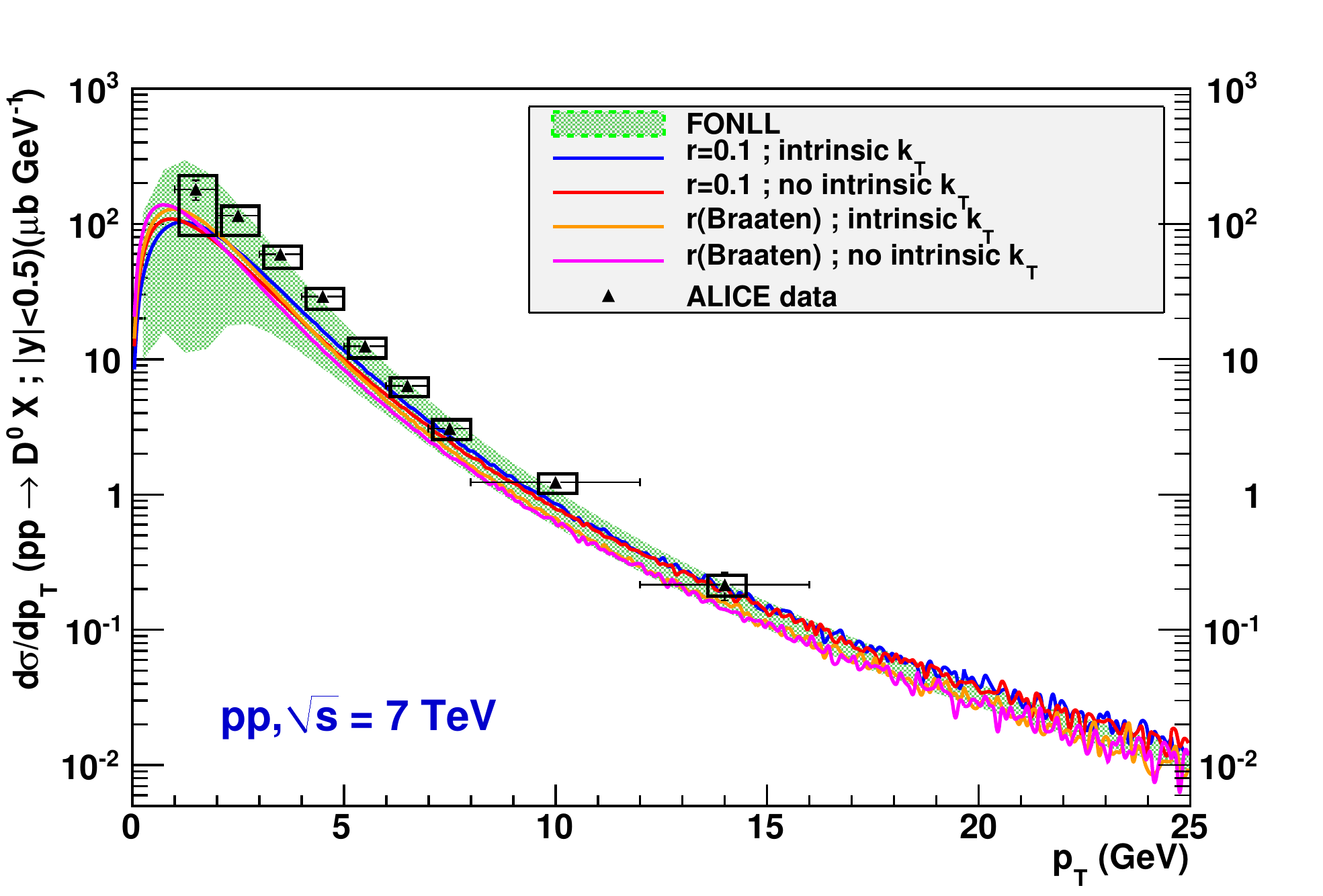}
\includegraphics[clip,width=0.48\textwidth]{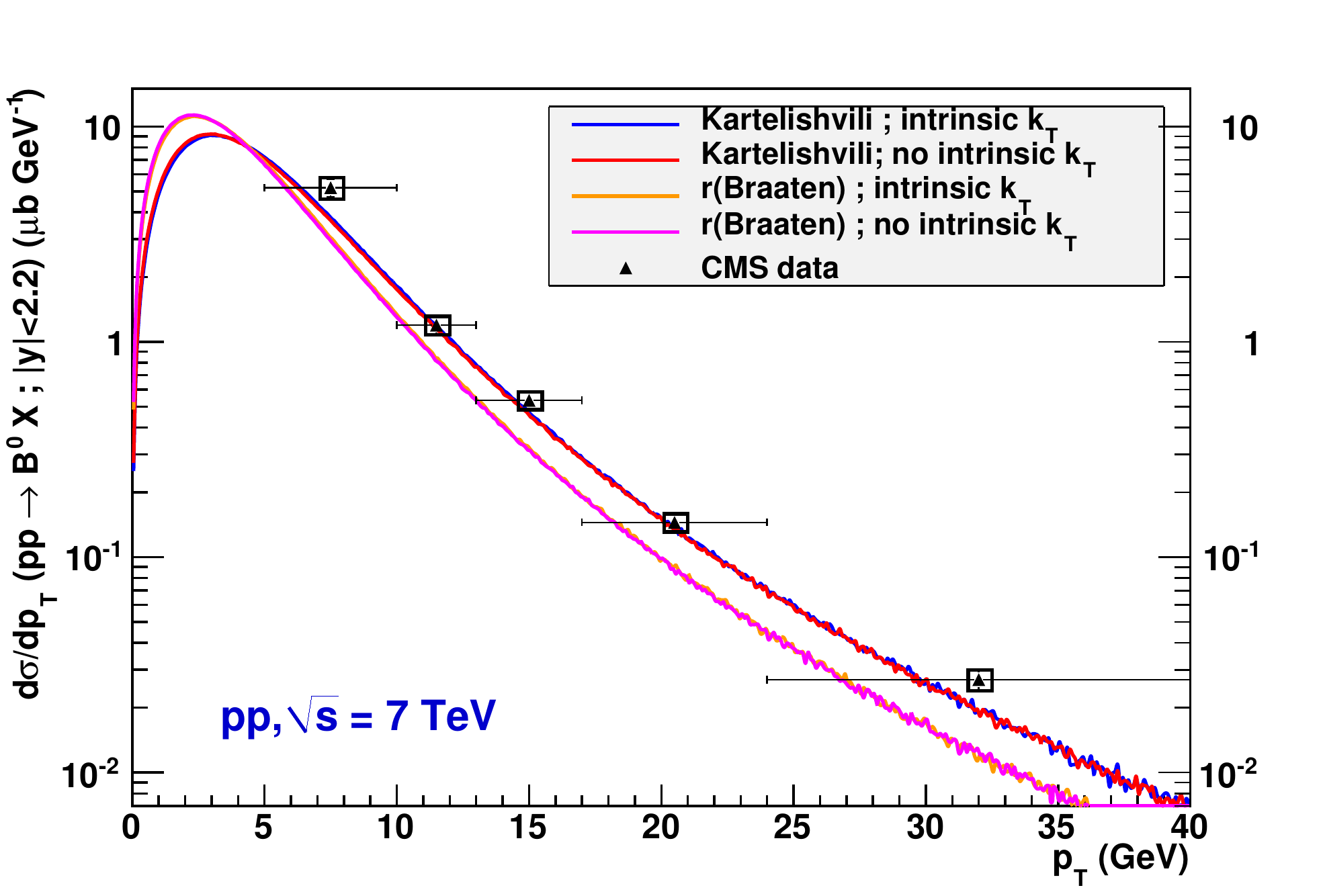}
\caption{$p_{T}$-differential inclusive cross-section in pp collisions 
at $\sqrt{s}=7$~TeV for $D^0$ mesons in $|y|<$0.5 (left) and for $B^0$ mesons 
in $|y|<$2.2 (right): predictions from POWHEG+PYTHIA under different 
assumptions on fragmentation functions and on partonic intrinsic-$k_T$ are 
compared with ALICE~\cite{ALICE_D_pp7TeV} and CMS~\cite{CMS_B0_pp7TeV} data. 
For charmed mesons, the theoretical uncertainty band of FONLL~\cite{FONLL_LHC} 
predictions is also superimposed.}
\label{fig:D0_B0_pp7TeV}
\end{center}
\end{figure}

As a test of the validity of our setup to make heavy-quark energy loss
studies at the LHC, we have checked the consistency of the charm and
bottom hadron spectra obtained from POWHEG (followed by the PYTHIA parton 
shower and by the heavy-quark fragmentation tools described above) with 
those measured at the LHC in pp collisions at $\sqrt{s}$=7~TeV. Results 
of these comparisons are shown in Fig.~\ref{fig:D0_B0_pp7TeV}.

In the left panel of Fig.~\ref{fig:D0_B0_pp7TeV}, 
the predictions for the $D^0$ meson inclusive $p_{T}$ distributions 
at central rapidity in pp collisions at $\sqrt{s}=7$~TeV are 
compared to ALICE data~\cite{ALICE_D_pp7TeV} and to the overall 
uncertainty band of the FONLL~\cite{FONLL_LHC} prediction, including 
also systematics from quark mass and $\mu$ scale variations.
Our results, obtained with different choices of the $r$ parameter of
the FF calculated in~\cite{Braaten} and with/without intrinsic-$k_T$, 
show a small systematic uncertainty from these variations, and are
consistent with the FONLL central prediction. The ALICE data points 
appear located on the upper edge of the FONLL uncertainty band (as it 
occurred already with D meson data at the Tevatron), and we can
conclude that within the large theoretical and experimental 
systematics they are also in rather good agreement with our 
calculations. A similar level of agreement with the ALICE data is
observed for the $p_{T}$ distributions of $D^+$ and $D^{*+}$ 
mesons, and also for those measured, with low statistics, at 
$\sqrt{s}$=2.76~TeV~\cite{ALICE_D_pp2.76TeV}.

In addition, in the right panel of Fig.~\ref{fig:D0_B0_pp7TeV}, our predictions
of the $B^0$ meson inclusive $p_{T}$ distributions at central
rapidity in pp collisions at $\sqrt{s}=7$~TeV are compared to CMS data 
from ref.~\cite{CMS_B0_pp7TeV}. A good agreement is observed, within 
respective uncertainties, mostly when Kartelishvili et al.~\cite{KAR}
parameterizations are used. 

\begin{figure}[htb]
\begin{center}
\includegraphics[clip,width=0.48\textwidth]{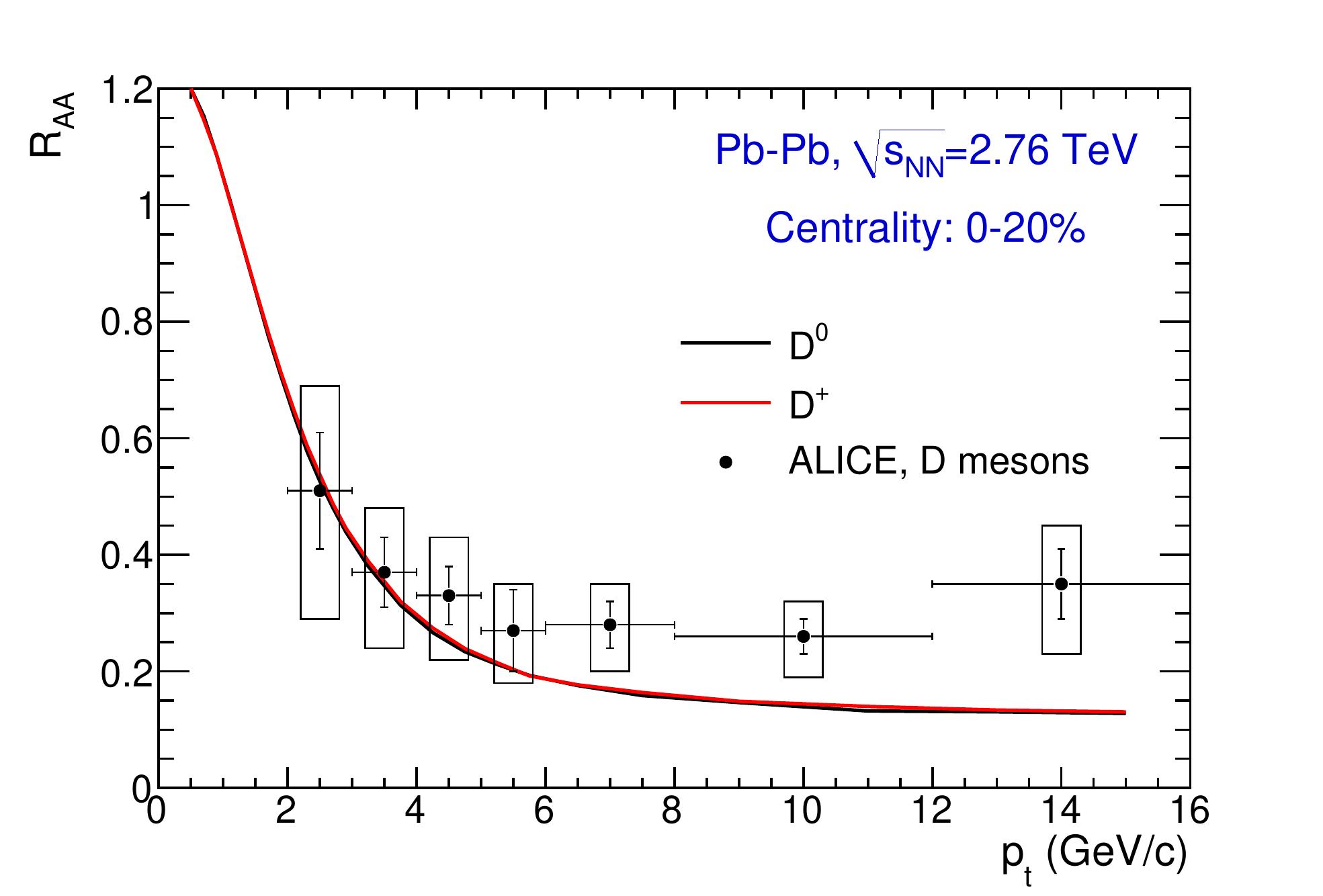}
\includegraphics[clip,width=0.48\textwidth]{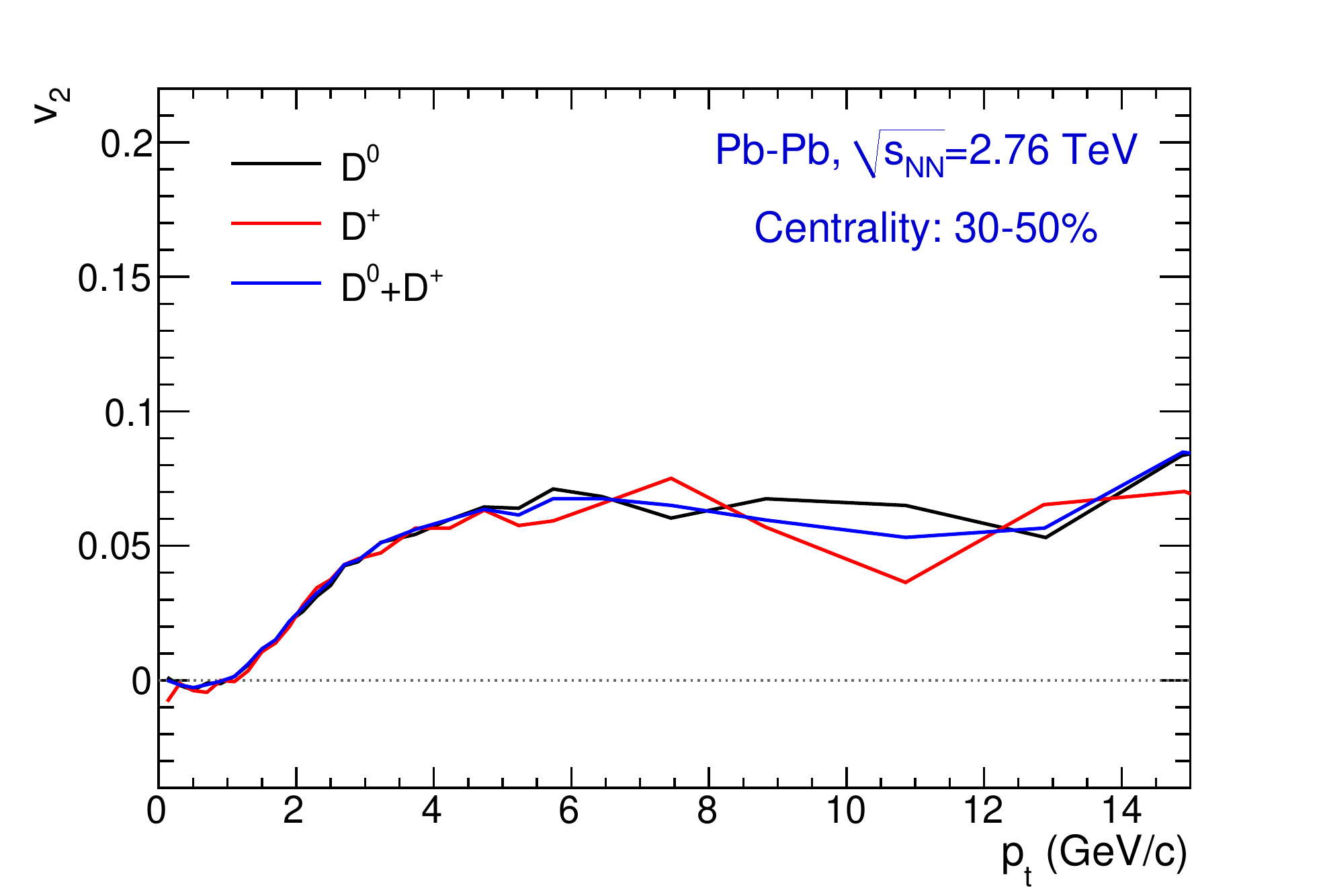}
\caption{(left) $R_{AA}(p_T)$ of prompt $D^0$ and $D^+$ mesons produced 
in Pb--Pb collisions at $\sqrt{s_{NN}}$=2.76~TeV at mid-rapidity ($|y|<$0.5) 
in the $0-20\%$ centrality class; predictions with the POWLANG model  
are compared with ALICE data~\cite{ALICE_D_RAA}  on average $R_{AA}$ 
of prompt D mesons versus $p_{T}$; (right) elliptic-flow $v_{2}$($p_{T}$) of 
prompt $D^0$ and $D^+$ mesons at mid-rapidity ($|y|<$0.5) in the $0-20\%$ 
centrality class.} 
\label{fig:RAAv2_D_ALICE}
\end{center}
\end{figure}

The effects of the Langevin evolution on heavy-quark spectra, as modeled in POWLANG, 
are displayed in the left panel of Fig.~\ref{fig:RAAv2_D_ALICE}, where our results
for the nuclear modification factor $R_{AA}$ of $D^0$ and  $D^+$ mesons produced 
at mid-rapidity in Pb--Pb collisions at $\sqrt{s_{NN}}$=2.76~TeV, in the $0-20\%$ 
centrality class, are compared to the ALICE data~\cite{ALICE_D_RAA}. The right panel 
of the same figure shows our predictions for the elliptic-flow $v_{2}$ of $D^0$ 
and  $D^+$ mesons produced at mid-rapidity in Pb--Pb collisions at 
$\sqrt{s_{NN}}$=2.76~TeV in the $30-50\%$ centrality class, that can be compared with 
the preliminary ALICE data shown at this conference in~\cite{ALICE_HP2012_D_v2},
but not yet published. 

Our predictions for $R_{AA}$ at low $p_{T}$  are in rather good agreement with the ALICE 
data, within the experimental and theoretical uncertainties, while the increasing trend 
of $R_{AA}$ measured at high $p_{T}$ is not reproduced. In addition, POWLANG model 
cannot reproduce the rather high $v_{2}$ values (up to 0.15-0.2) measured by 
ALICE~\cite{ALICE_HP2012_D_v2} at low $p_{T}$.

\begin{figure}[t]
\begin{center}
\includegraphics[clip,width=0.48\textwidth]{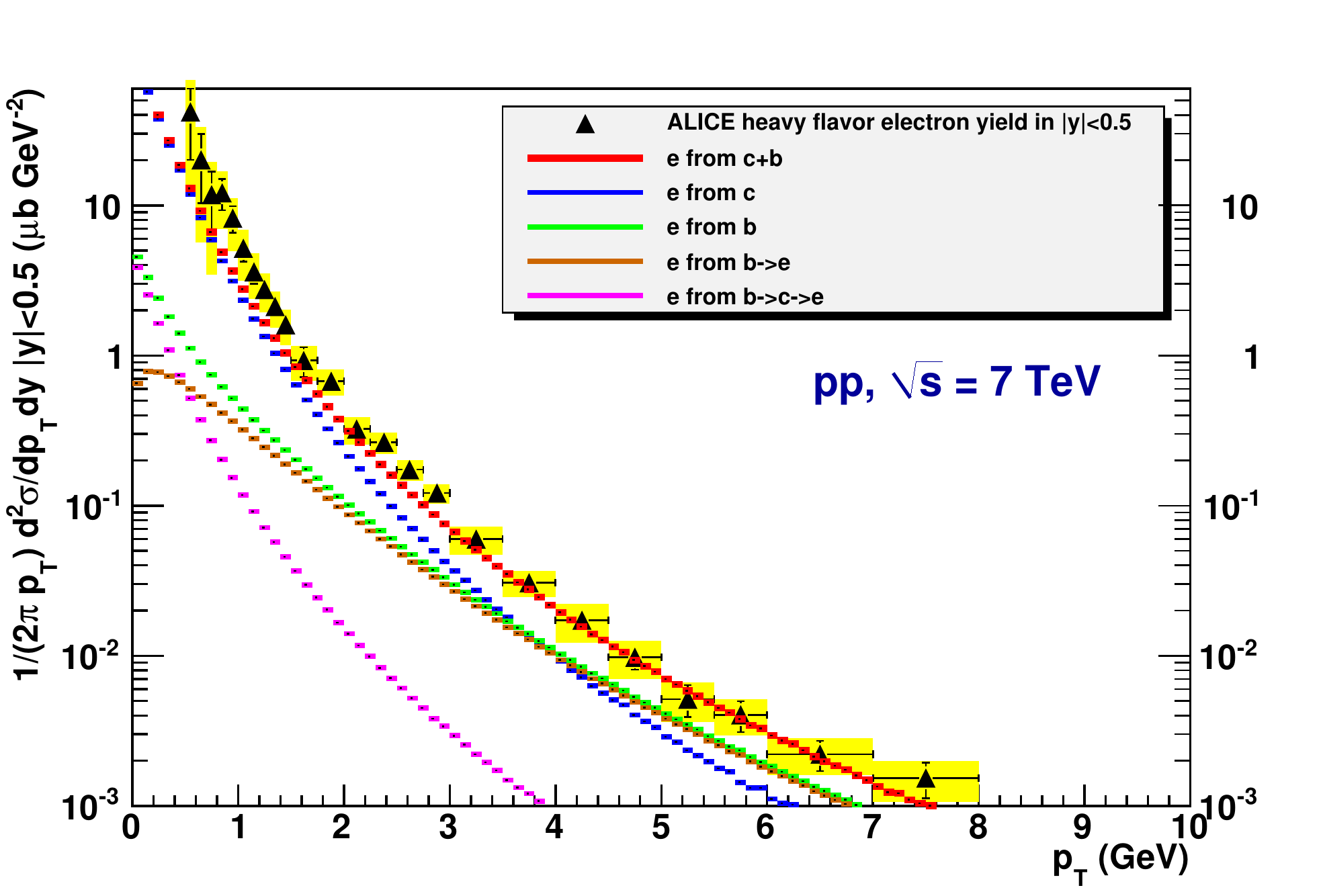}
\includegraphics[clip,width=0.48\textwidth]{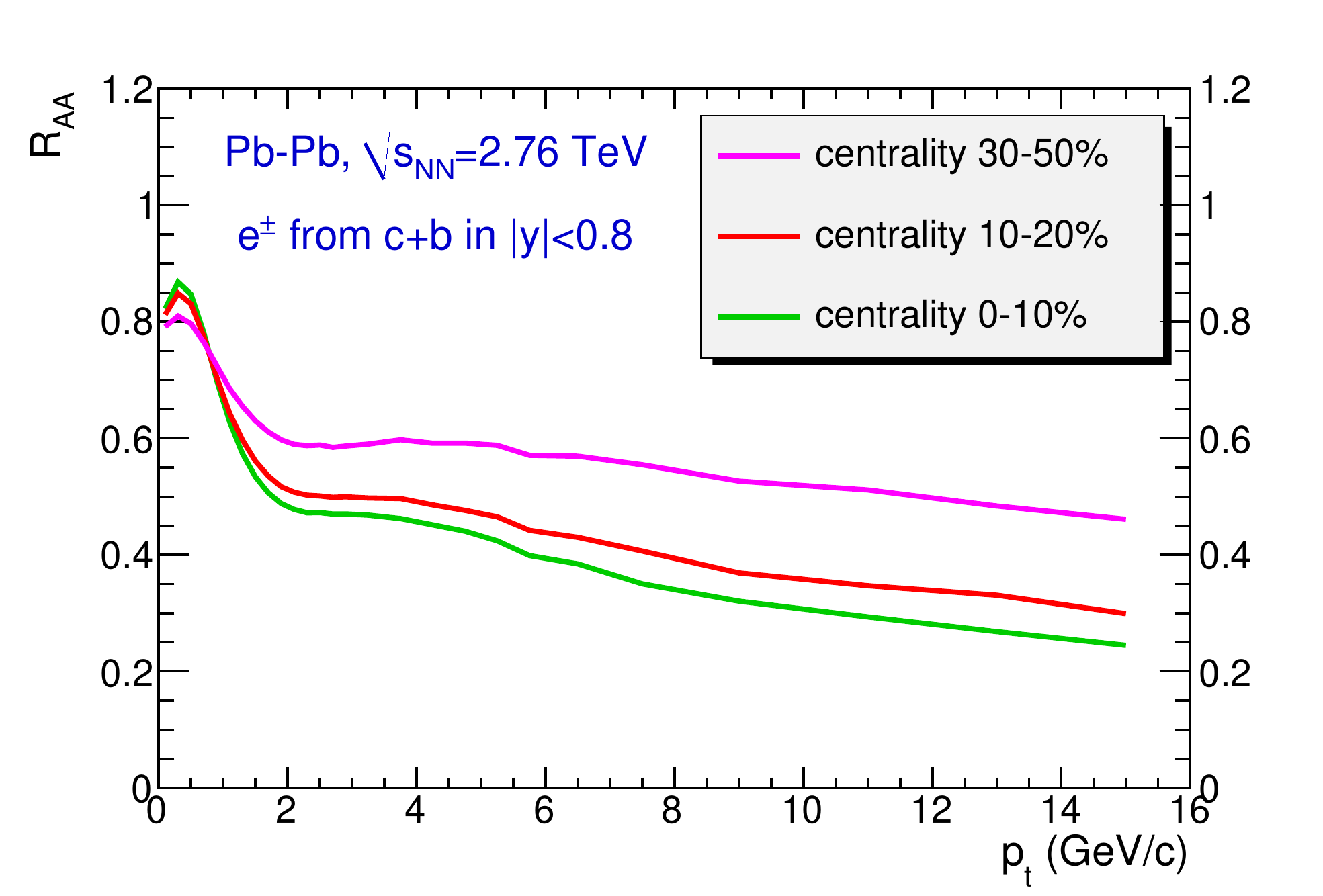}
\caption{(left) invariant $p_{T}$-differential inclusive cross-section 
of electrons from heavy-flavor decays in pp collisions at $\sqrt{s}=7$~TeV 
at central rapidity ($|y|<$0.5); ALICE~\cite{ALICE_HFE_pp} data on the total 
inclusive electron yield are compared to our predictions with POWHEG+PYTHIA, 
including also the separate contributions to the electron yield from charm 
and beauty semi-leptonic decays; (right) $R_{AA}(p_T)$ of inclusive heavy-flavor
electrons in Pb--Pb collisions at $\sqrt{s_{NN}}$=2.76~TeV at mid-rapidity 
($|y|<$0.8) and for three centrality classes: $0-10\%$, $0-20\%$ and $30-50\%$.} 
\label{fig:HFE_pp_RAA}
\end{center}
\end{figure}

To conclude, we show our predictions for the measurement of
semi-leptonic heavy-flavor decays in ALICE, at cental rapidity. 
In Fig.~\ref{fig:HFE_pp_RAA} (left panel) our prediction of the
differential invariant production cross section at central rapidity of electrons
from heavy-flavor decays in pp collisions at $\sqrt{s}$=7~TeV 
is compared to ALICE data~\cite{ALICE_HFE_pp}. Separate contributions
to electron yields from different decay channels ($D\!\to\! e$,
$B\!\to\! e$ or $B\!\to\! D \to e$) are also shown. 
As in the case of D mesons we can conclude that, within the
experimental uncertainties and the systematics affecting our
calculations, our prediction for the heavy-flavor electron spectrum 
in pp collisions agrees reasonably with the ALICE data.

Finally, in the right panel of Fig.~\ref{fig:HFE_pp_RAA}, we show our predictions 
for the $R_{AA}$ of the total inclusive electron yields from heavy-flavor decays
at central rapidity in Pb--Pb collisions at $\sqrt{s_{NN}}$=2.76~TeV,  
in three different centrality classes ($0-10\%$, $10-20\%$ and
$30-50\%$). These predictions can be compared to the preliminary 
data reported by ALICE by~\cite{ALICE_HP2012_HFE_RAA}, and 
at least qualitatively, before official data publication, we can already conclude 
that heavy-flavor electron $R_{AA}$ measured by ALICE are reasonably described 
by POWLANG at moderate $p_{T}$. 

\section{Summary}

The POWLANG setup provides a reasonable description of the quenching of heavy-flavor spectra 
measured by ALICE at the LHC in Pb--Pb collisions at $\sqrt{s_{NN}}$=2.76~TeV, for moderate values
of $p_{T}$. In the same $p_{T}$ range the experimental results for $v_{2}$ are underpredicted. This 
could either point to a shortcoming of a perturbative picture of rescattering in the hot medium, 
or to the contribution at hadronization of coalescence, whose implementation is left for future 
investigations. 

On the other hand at high $p_{T}$ the Langevin approach seems to overestimate the observed amount 
of quenching.





\bibliographystyle{elsarticle-num}


\begin{thebibliography}{99}

 
\bibitem{Langevin2011} W. M. Alberico et al., Eur. Phys. J. C 54 (2011) 1666.

\bibitem{QM2011} M. Monteno et al., J. Phys. G: Nucl. Part. Phys. 38 (2011) 124144.


\bibitem{POWHEG-hvq} S. Frixione, P. Nason and G. Ridolfi, JHEP 09 (2007) 126. 

\bibitem{Pythia} T. Sjostrand, S. Mrenna and P. Z. Skands, JHEP 05 (2006) 026.

\bibitem{EPS09} K. J. Eskola, H. Paukkunen and C. A. Salgado, JHEP 04 (2009) 065.


\bibitem{kolb} P. F. Kolb, J. Sollfrank and U. Heinz, Phys. Rev. C 62 (2000) 054909.

\bibitem{romatschke} P. Romatschke and U. Romatschke, Phys. Rev. Lett. 99 (2007) 172301. 


\bibitem{ZEUS} S. Chekanov et al., ZEUS Collaboration,  Eur. Phys. J. C 44 (2005) 351.

\bibitem{ALEPH1} R. Barate et al., ALEPH Collaboration, Eur. Phys. J. C 16 (2000) 597-611, arXiv:hep-ex/9909032.


\bibitem{HFAG} D. Asner et al., Heavy Flavour Averaging Group, arXiv:1010.1589 [hep-ex].


\bibitem{Braaten} E. Braaten, K.M. Cheung, S. Fleming and T. C. Yuan, Phys. Rev. D 51 (1995) 4819.

 
\bibitem{FONLL_D_Tevatron} M. Cacciari and P. Nason, JHEP 09 (2003) 006, arXiv:hep-ph/0306212.

\bibitem{KAR} V. G. Kartvelishvili, A. K. Likhoded, V. A. Petrov, Phys. Lett. B78 (1978) 615.

\bibitem{FONLL_B_Tevatron} M. Cacciari et al., JHEP 07 (2004) 033. 

\bibitem{ALEPH2} A. Heister, et al., ALEPH Collaboration, Phys. Lett. B 512 (2001) 30-48.

\bibitem{SLD} K. Abe et al., SLD Collaboration, Phys. Rev. D 65 (2002) 092006 [Erratum-ibid. D66 (2002) 079905].


\bibitem{PDG2012} J. Beringer et al., Particle Data Group, Phys. Rev. D 86 (2012) 010001.


\bibitem{ALICE_D_pp7TeV} B. Abelev et al., ALICE Collaboration, JHEP 01 (2012) 128, arXiv:1111.1553v2 [hep-ex].


\bibitem{FONLL_LHC} M. Cacciari et al., arXiv:1205.6344 [hep-ph]. 


\bibitem{ALICE_D_pp2.76TeV}  B. Abelev et al., ALICE Collaboration, arXiv:1205.1407 [hep-ex]. 


\bibitem{CMS_B0_pp7TeV} S. Chatrchyan, et al., CMS Collaboration, Phys. Rev. Lett. 106 (2011) 252001.


\bibitem{ALICE_D_RAA} B. Abelev et al., ALICE Collaboration, arXiv:1203.2160 [hep-ex].

\bibitem{ALICE_HP2012_D_v2} Giacomo Ortona, for the ALICE Collaboration, in {\it Proceedings of Hard Probes 2012, 27 May - 1 June 2012, Cagliari (Italy)}. 


\bibitem{ALICE_HFE_pp} B. Abelev et al., ALICE Collaboration, arXiv:1205.5423 [hep-ex].


\bibitem{ALICE_HP2012_HFE_RAA} M.J. Kweon, for the ALICE Collaboration, in {\it Proceedings of Hard Probes 2012, 27 May - 1 June 2012, Cagliari (Italy)}.   



\end{thebibliography}



\end{document}